\documentclass[conference,9pt]{IEEEtran}
\IEEEoverridecommandlockouts

\usepackage{cite}
\usepackage{amsmath,amssymb,amsfonts}
\usepackage{algorithmic}
\usepackage{graphicx}
\usepackage{textcomp}
\usepackage[table]{xcolor}

\usepackage{booktabs}
\usepackage[hidelinks]{hyperref}
\usepackage{siunitx}
\usepackage{mathtools}
\usepackage{comment}
\usepackage{multirow}
\usepackage{makecell}
\usepackage{arydshln}   
\usepackage{enumitem}
\usepackage[ruled,linesnumbered,vlined]{algorithm2e}
\usepackage{inconsolata}
\usepackage{tabularx}   
\usepackage{array}  

\usepackage{tikz}
\usetikzlibrary{calc}
\usetikzlibrary{patterns,patterns.meta}
\usetikzlibrary{arrows.meta}
\usepackage[outline]{contour}
\contourlength{1.3pt}

\makeatletter
\let\MYcaption\@makecaption
\makeatother
\usepackage[font=footnotesize,subrefformat=parens]{subcaption}
\makeatletter
\let\@makecaption\MYcaption
\makeatother

\usepackage{xspace}
\makeatletter
\DeclareRobustCommand\onedot{\futurelet\@let@token\@onedot}
\def\@onedot{\ifx\@let@token.\else.\null\fi\xspace}

\makeatother

\def\equationautorefname~#1\null{(#1\null)}

\renewcommand{\sectionautorefname}{Section}
\renewcommand{\subsectionautorefname}{\sectionautorefname}

\let\orgautoref\autoref
\providecommand{\Autoref}[1]
{%
\def\figureautorefname{Figure}%
\def\subfigureautorefname{Figure}%
\orgautoref{#1}%
}
\renewcommand{\autoref}[1]
{%
\def\figureautorefname{Fig.}%
\def\subfigureautorefname{\figureautorefname}%
\def\sectionautorefname{Sec.}%
\def\subsectionautorefname{\sectionautorefname}%
\def\subsectionautorefname{\sectionautorefname}%
\orgautoref{#1}%
}

\newcommand{\vect}[1]{\mbox{\boldmath $#1$}}

\newcommand{\norm}[1]{\left\lVert#1\right\rVert}

\newcommand{\trans}[1]{#1^\mathsf{T}}

\DeclareMathOperator*{\argmin}{arg\,min}\def\appendixautorefname~#1\null{~#1 \null}

\DeclareFontEncoding{LS1}{}{}
\DeclareFontSubstitution{LS1}{stix}{m}{n}
\DeclareSymbolFont{stixletters}{LS1}{stix}{m}{it}
\DeclareMathAccent{\anticausal}{\mathord}{stixletters}{"91}
\DeclareMathAccent{\causal}{\mathord}{stixletters}{"92}
\DeclareMathAccent{\bidirectional}{\mathord}{stixletters}{"95}
\DeclareMathAccent{\noncausal}{\mathord}{stixletters}{"95}

\makeatletter
\newcommand{\figcaption}[1]{\def\@captype{figure}\caption{#1}}
\newcommand{\tblcaption}[1]{\def\@captype{table}\caption{#1}}
\makeatother

\makeatletter 
\newcommand{\linebreakand}{%
  \end{@IEEEauthorhalign}
  \hfill\mbox{}\par
  \mbox{}\hfill\begin{@IEEEauthorhalign}
}
\makeatother 

\def\BibTeX{{\rm B\kern-.05em{\sc i\kern-.025em b}\kern-.08em
    T\kern-.1667em\lower.7ex\hbox{E}\kern-.125emX}}

\newcommand{\customdashline}[1]{%
  \noalign{\vskip\aboverulesep} 
  \cdashline{#1}[1pt/1pt]
  \noalign{\vskip\belowrulesep} 
}

\makeatletter
\patchcmd{\@algocf@start}
  {-1.5em}
  {0pt}
{}{}
\makeatother
\begin{document}
\abovedisplayskip=2pt
\belowdisplayskip=\abovedisplayskip

\setlength\textfloatsep{8pt}
\setlength\dbltextfloatsep{8pt}
\setlength\floatsep{8pt}
\setlength\dblfloatsep{8pt}
\captionsetup[figure]{skip=6pt}
\captionsetup[table]{skip=-1pt}
\subcaptionsetup[figure]{skip=2pt}

\title{Over-Tightening-Aware Pseudo-Labeling for\\Tight-Boundary Speaker Diarization}
\author{\IEEEauthorblockN{Shota Horiguchi, Takanori Ashihara, Marc Delcroix, Naohiro Tawara, Alexis Plaquet}
\IEEEauthorblockA{NTT, Inc., Japan}}

\maketitle

\begin{abstract}
Training speaker diarization models on loose labels, such as speech segments with padded boundaries or filled pauses, often results in similarly loose model outputs.
To obtain tighter boundaries, pseudo-labeling based on the averaged outputs of causal and anticausal models has been proposed.
However, since the pseudo-labels are estimation-based, they can suffer from over-tightening, which increases missed detections that can propagate as unrecoverable errors to downstream tasks.
This paper carefully analyzes the causes of over-tightening and proposes three approaches to address them: (i) removing pause filling rather than padding, (ii) introducing a burn-in phase to mitigate missed detections near the beginning of causal and anticausal predictions, and (iii) making pseudo-label-based co-training aware of the non-causal model used for final inference.
Experimental results show that the proposed method reduces missed detections caused by over-tightening and improves both diarization accuracy and downstream multi-talker ASR performance.
\end{abstract}


\section{Introduction}\label{sec:introduction}
Speaker diarization, which estimates speaker-wise speech segments from multi-speaker audio, has become increasingly important.
Recent multi-speaker ASR systems based on neural networks often adopt architectures that generate transcriptions based on diarization results~\cite{raj2022speaker,lin2025diarization,park2025sortformer,polok2026dicow,polok2026sedicow,li2026dm}.
Speaker diarization also plays an important role in conversational analysis and in constructing training data for spoken dialogue models~\cite{defossez2024moshi,fu2024investigating,ohashi2025towards,veluri2024beyond,cui2026turnguide,jung2026sommelier}.

Neural networks that directly estimate speaker-wise speech segments from an input recording, either an entire session or a chunked input, are referred to as end-to-end neural diarization (EEND) models~\cite{fujita2019end1,fujita2019end2,medennikov2020targetspeaker,kinoshita2021integrating,horiguchi2022encoderdecoder,wang2023target,harkonen2024eend,cheng2025sequence}.
Such models require a large amount of training data, and a common recent approach is to combine multiple corpora for training~\cite{bredin2023pyannote,plaquet2023powerset,han2025leveraging,han2025fine}.
However, it has been pointed out that the tightness of annotations varies substantially across corpora~\cite{horiguchi2025can}.
Corpora specifically designed for diarization tend to annotate only intervals where speech is actually present as speech regions.
In contrast, corpora designed for ASR may include non-speech regions within speech segments due to onset/offset padding and the avoidance of segment splits at short pauses to preserve semantic continuity.
When corpora with different annotation criteria are mixed for training, the model may learn to imitate corpus-dependent loose or tight segment boundaries, making its behavior less predictable, especially on out-of-domain data~\cite{horiguchi2025can}.
To avoid this issue, it is important to make the tightness of all annotations as consistent as possible during training.
Since loose prediction can be relatively easily recovered from tight prediction using operations such as morphological closing~\cite{horiguchi2020utterance,boeddeker2023multi,boeddeker2024tssep}, unifying annotations into tight labels is considered advantageous.

The simplest approach is to re-annotate tight speech segments by applying forced alignment to speaker-separated recordings using oracle transcripts~\cite{watanabe2020chime,horiguchi2025can,polok2026mind}.
However, this approach is applicable only when such recordings are available, and their transcripts are provided.
A more recent study proposed exploiting causal--anticausal consistency to generate tight pseudo-labels without the need for forced alignments~\cite{horiguchi2026tight}.
This method relies on the following properties: (i) even when trained on loose segments, a causal model cannot perform onset padding because it cannot access future information, (ii) an anticausal model cannot perform offset padding because it cannot access past information, and (iii) neither model can perfectly reproduce pause filling.
Therefore, combining the outputs of causal and anticausal models enables the generation of tight pseudo-labels.
These pseudo-labels are then used to train a non-causal model, enabling tight-label training without requiring large-scale tight annotations in the corpus or the above-mentioned resources.
To further enhance this effect, the causal and anticausal models are co-trained, with their parameters updated at each training iteration using the pseudo-labels generated by the current models.
However, these labels remain prediction-based and can still become overly tight, particularly during co-training, even with the several techniques introduced in the method.
Since a model can learn over-tightened behavior in the same way that it can learn looseness, this problem leads to increased missed detections in the final predictions.
Such missed detections propagate errors to downstream tasks, such as ASR.

Motivated by this issue, this paper extensively analyzes the factors that cause over-tightening in the pseudo-labeling process described above.
Based on this analysis, we propose methods to address these factors, aiming to improve both diarization and downstream ASR performance.
Specifically, the proposed methods are threefold: (i) since looseness is primarily caused by pause filling rather than boundary padding, we introduce a tightening method that focuses on mitigating pause filling while preserving loose segment boundaries; (ii) since the causal and anticausal models used for pseudo-label generation tend to produce missed detections near the beginning of their respective sequences, we introduce a burn-in phase by providing preceding context to the causal model and the following context to the anticausal model; and (iii) since the pseudo-labels generated by the best-performing causal and anticausal models do not necessarily yield the best non-causal model, we incorporate the non-causal model into the co-training framework.
These techniques are simple, yet they substantially improve tight-boundary speaker diarization.
Compared with the conventional method, the proposed method reduced the DER gap to the ideal-tight-label topline by about \SI{40}{\percent} on AMI and \SI{60}{\percent} on AliMeeting.
The ASR systems using the resulting diarization outputs also achieved performance close to that obtained using diarization outputs from the model trained with tight labels.

\section{Tight-Boundary Speaker Diarization: A Review}\label{sec:tight_boundary}
\subsection{Problem formulation}
In most end-to-end speaker diarization frameworks, speaker diarization is formulated as a fully supervised classification problem.
Let $X\in\mathbb{R}^{F\times L}$ be an input sequence of length $L$, where each time step has $F$ dimensions. Here, $F=1$ corresponds to a raw waveform and $F>1$ corresponds to arbitrary spectral features.
A diarization model based on a typically non-causal neural network $f(\cdot)$ transforms the input into frame-wise posterior probabilities as
\begin{equation}
    Q\coloneqq\left[\vect{q}_1,\dots,\vect{q}_T\right]=f(X)\in(0,1)^{C\times T},
\end{equation}
where $\vect{q}_t\coloneqq\trans{[q_{1,t},\dots,q_{C,t}]}\in(0,1)^C$ represents posteriors for $C$ classes at $t$-th frame with $1\leq t\leq T$.
In this study, we use a powerset classification formulation, where each class represents a possible set of active speakers~\cite{plaquet2023powerset}.
Thus, $C=\sum_{m=0}^M\binom{S}{m}$, where $S$ is the maximum number of speakers per input and $M$ is the maximum number of speakers who speak simultaneously at each frame.
These powerset-based posteriors $Q$ can be transformed to the speaker-wise posteriors $P\coloneqq[\vect{p}_1,\dots,\vect{p}_T]\in(0,1)^{S\times T}$ by using the function $\mathsf{Q2P}:Q\mapsto P$ defined as
\begin{equation}
    p_{s,t}\coloneqq\sum_{c:s\in\mathcal{S}_c}q_{c,t},\label{eq:q2p}
\end{equation}
where $\vect{p}_t\coloneqq\trans{[p_{1,t},\dots,p_{S,t}]}$, and $\mathcal{S}_c$ is the set of speakers corresponding to powerset class $c$.

The model $f(\cdot)$ is optimized using the cross-entropy loss.
Since the output speaker order is arbitrary, the speaker order of the model output must be aligned with that of the reference label beforehand:
\begin{equation}
P^{(Y)}=\argmin_{P'\in\Pi(P)}\norm{P'-Y}_F,\label{eq:perm}
\end{equation}
where $Y\coloneqq(y_{s,t})\in\{0,1\}^{S\times T}$ denotes the reference label, with $y_{s,t}=1$ if speaker $s$ is speaking at frame $t$ and $0$ otherwise, $\Pi(P)$ denotes the set of all possible speaker-permuted $P$, and $\norm{\cdot}_F$ denotes the Frobenius norm.
The speaker-aligned powerset posteriors $q^{(Y)}_{c,t}$ can be obtained accordingly using the optimal permutation computed in \autoref{eq:perm}.
Then the loss to be minimized is computed as
\begin{equation}
\ell(Q,Y)=-\sum_{c=1}^{C}\sum_{t=1}^T z_{c,t}\log q_{c,t}^{(Y)},\label{eq:loss}
\end{equation}
where $z_{c,t}$ is the following powerset-based reference label:
\begin{equation}
    z_{c,t}=\begin{cases}
        1 & (\text{if } \{s\mid y_{s,t}=1\}=\mathcal{S}_c),\\
        0 & (\text{otherwise}).
    \end{cases}
\end{equation}

Most end-to-end neural diarization studies treated annotated labels $\tilde{Y}\coloneqq(\tilde{y}_{s,t})\in\{0,1\}^{S\times T}$ as the reference, i.e., $\tilde{Y}=Y$.
However, the original annotations in some corpora are not sufficiently tight for speaker diarization, mainly because they were developed for ASR and thus avoid excessive segmentation that would break semantic continuity.
If a model is trained using such loose annotations, it tends to produce similarly loose speech segments~\cite{horiguchi2025can}.
Motivated by this issue, tight-boundary speaker diarization has recently been proposed as a framework for training a diarization model using only loose annotations while enabling it to predict tight speech boundaries~\cite{horiguchi2026tight}.
It assumes one-sided label noise in the speech class, like
\begin{equation}
y_{s,t}\in\begin{cases}
\left\{0,1\right\} & (\tilde{y}_{s,t}=1),\\
\left\{0\right\} & (\tilde{y}_{s,t}=0).
\end{cases}
\end{equation}
Such noise arises from onset or offset padding and pause filling.

\subsection{Pseudo-labeling via causal--anticausal consistency}\label{sec:pseudo_labeling}
In tight-boundary speaker diarization, the loss in \autoref{eq:loss} cannot be computed directly because the label $y_{s,t}$ required for computing the loss is unknown.
To address this issue, as explained in \autoref{sec:introduction}, previous work exploits the properties that causal and anticausal models cannot perform onset padding and offset padding, respectively, and that neither model can perform pause filling perfectly~\cite{horiguchi2026tight}.
Specifically, tight pseudo-labels are generated by combining the outputs of causal and anticausal models trained on loose annotations, which are used to compute the loss.
The overall procedure for training a non-causal model that can output tight predictions is summarized below.
\begin{itemize}
\item \textbf{Step 1:} Train causal and anticausal models independently using the original loose labels $\tilde{Y}$.
\item \textbf{Step 2:} Co-train causal and anticausal models from the pretrained weights. At each training iteration, pseudo-labels are generated on the fly from the outputs of the causal and anticausal models, and both models are updated using these pseudo-labels.
\item \textbf{Step 3:} Train a non-causal model using the final pseudo-labels.
\end{itemize}
\begin{algorithm}[t]
    \caption{Pseudo-label generation using causal and anticausal prediction \cite{horiguchi2026tight}}\label{algo:pseudo}
    \DontPrintSemicolon
    \SetKwInOut{Input}{Input}
    \SetKwInOut{Output}{Output}
    \SetKwComment{Comment}{$\triangleright$\ }{}
    \Input{Causal prediction $\causal{Q}\in(0,1)^{C\times T}$, anticausal prediction $\anticausal{Q}\in(0,1)^{C\times T}$, loose label $\tilde{Y}\in\{0,1\}^{S\times T}$}
    \Output{Tightened pseudo-label $Y_\mathrm{tight}\in\{0,1\}^{S\times T}$}
    \BlankLine
    
    $\causal{P}=\mathsf{Q2P}(\causal{Q}),\;\anticausal{P}=\mathsf{Q2P}(\anticausal{Q})$\label{algline:q2p}\Comment*[r]{Powerset to multilabel}
    Align the speaker permutation of $\causal{P}$ and $\anticausal{P}$ to $\tilde{Y}$\label{algline:align}\;
    Mitigate speaker confusion by swapping posteriors of missed and false alarm speakers in $\causal{P}$ and $\anticausal{P}$, by referring $\tilde{Y}$\label{algline:swap}\;
    $Y_\mathrm{tight}=\tilde{Y}\odot\phi_{\tau}\bigl(\frac{\causal{P}+\anticausal{P}}{2}\bigr)$\label{algline:mask}\;
    Restore over-tightened segments in $Y_\mathrm{tight}$\label{algline:restore}
\end{algorithm}

For later discussion, we describe the pseudo-label generation process, which is used in \textbf{Step 2} and \textbf{Step 3}, in more detail using \autoref{algo:pseudo}.
First, the posterior outputs of the causal and anticausal models over powerset classes, $\causal{Q}$ and $\anticausal{Q}$, are converted into speaker-wise posteriors (\autoref{algline:q2p}).
Next, since the outputs of these models are invariant to speaker permutation, they are aligned with $\tilde{Y}$ (\autoref{algline:align}).
These model outputs may contain speaker confusion, and using them directly as pseudo-labels can degrade performance.
To mitigate this effect, the prior study proposed two correction strategies: a conservative method based on voice activity detection (VAD) and a more aggressive method based on speaker counting (SC).
In this paper, we adopt the SC-based method, which refers to the loose label $\tilde{Y}$ and, at each frame, swaps the posteriors of missed speakers with those of falsely detected speakers (\autoref{algline:swap}).
Finally, we obtain the tightened pseudo-label $Y_\mathrm{tight}$ by first applying the element-wise thresholding function $\phi_\tau(\cdot)$ with threshold $\tau$ to the average of $\causal{P}$ and $\anticausal{P}$, and then using the resulting binary mask to mask the loose label $\tilde{Y}$ (\autoref{algline:mask}).
Finally, since the tightened labels are derived from model predictions, they may be overly tightened. Assuming that speech should not completely disappear within a loose segment, we restore such disappeared segments (\autoref{algline:restore}).
The prior study adopts a more relaxed criterion, restoring a loose segment only when more than \SI{50}{\percent} of the segment has disappeared.

The prior work pioneered tight-boundary speaker diarization by formulating it as a new problem and proposing the first solution~\cite{horiguchi2026tight}.
However, its method still has room for improvement.
One of its major drawbacks is over-tightening, which increases missed detections and harms downstream tasks.
This paper aims to improve this approach by reducing missed detection from multiple perspectives.

\subsection{Model architecture}\label{sec:architecture}
Large-scale pretrained models, like WavLM~\cite{chen2022wavlm} used in DiariZen~\cite{han2026efficient}, usually do not support causal or anticausal inference.
The prior study employs a multi-speaker embedding extractor as a pretrained temporal encoder instead, with the backend consisting of a bidirectional long short-term memory (LSTM) and a linear layer to obtain frame-wise posteriors~\cite{horiguchi2025pretraining}.
This paper follows the same strategy and uses the ReDimNet-B2-based architecture for the pretrained encoder~\cite{yakovlev2024reshape}.
For causal and anticausal models, each convolutional kernel is replaced with a causal and anticausal variant of the same kernel size, and self-attention modules in Transformer encoders are applied with causal or anticausal masks, respectively.
The bidirectional LSTM is also replaced with a unidirectional LSTM in the left-to-right and right-to-left directions, respectively.

\section{Proposed Methods}
This paper analyzes multiple factors that cause over-tightening in SC-based tightening.
Based on this analysis, we propose solutions to address each factor, aiming to improve the performance of both diarization and downstream tasks.

\subsection{Pause-filling-focused tightening}\label{sec:pause_fill}
\begin{figure}[t]
\subfloat[Pause filling\label{fig:pause_filling}]{%
\begin{minipage}[t]{\linewidth}
    \centering
    \begin{tikzpicture}[semithick,auto,
label/.style={
    draw=none,
    align=center,
    font=\footnotesize,
    inner sep=0,
    outer sep=0
},
]%

\definecolor{spkblue}{HTML}{0080B1}
\definecolor{spkred}{HTML}{E4002B}
\definecolor{spkgreen}{HTML}{06C755}

\draw[black, line width=0.5pt] (0em,-1.2em) -- (9.0em,-1.2em);
\path let \p1 = ($(0.5em,0em)$),
          \p2 = ($(8.5em,-1.2em)$) in
    [pattern={Lines[angle=45,distance=2pt]},pattern color=spkblue,draw=black] (\p1) rectangle (\p2);
\node[font=\scriptsize,text=black,inner sep=0pt,anchor=base] at ($(4.5em,-0.8em)$) {\contour{white}{Loose label}};

\draw[black, line width=0.5pt] (0em,-3.6em) -- (9.0em,-3.6em);
\path let \p1 = ($(1.0em,-2.4em)$),
          \p2 = ($(3.5em,-3.6em)$) in
    [pattern={Lines[angle=-45,distance=2pt]},pattern color=spkred,draw=black] (\p1) rectangle (\p2);
\node[font=\scriptsize,text=black,inner sep=0pt,anchor=base] at ($(2.25em,-3.2em)$) {\contour{white}{Tight}};
\path let \p1 = ($(5.5em,-2.4em)$),
          \p2 = ($(8.0em,-3.6em)$) in
    [pattern={Lines[angle=-45,distance=2pt]},pattern color=spkred,draw=black] (\p1) rectangle (\p2);
\node[font=\scriptsize,text=black,inner sep=0pt,anchor=base] at ($(6.75em,-3.2em)$) {\contour{white}{label}};

\draw[black,dash pattern=on 2pt off 1pt,line width=0.5pt,color=gray] (3.5em,-3.6em) -- (3.5em,-4.5em);
\draw[black,dash pattern=on 2pt off 1pt,line width=0.5pt,color=gray] (5.5em,-3.6em) -- (5.5em,-4.5em);
\draw[<->, >={Latex[length=3pt,width=3pt]}, black, line width=0.4pt]
  (3.5em,-4.2em) -- node[midway, below, font=\scriptsize, inner sep=4pt] {$d_\text{pause}$}
  (5.5em,-4.2em);

\end{tikzpicture}%
\hfill
    \includegraphics[width=0.32\linewidth]{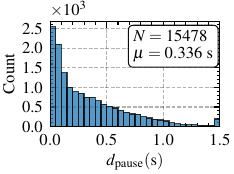}\hfill
    \includegraphics[width=0.32\linewidth]{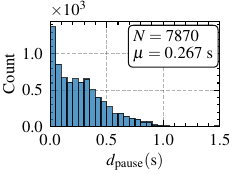}
\end{minipage}%
}
\hfill
\subfloat[Onset boundary shift\label{fig:onset_padding}]{%
\begin{minipage}[t]{\linewidth}
    \centering
    \begin{tikzpicture}[semithick,auto,
label/.style={
    draw=none,
    align=center,
    font=\footnotesize,
    inner sep=0,
    outer sep=0
},
]%

\definecolor{spkblue}{HTML}{0080B1}
\definecolor{spkred}{HTML}{E4002B}
\definecolor{spkgreen}{HTML}{06C755}

\draw[black, line width=0.5pt] (0em,-1.2em) -- (9.0em,-1.2em);
\path let \p1 = ($(1em,0em)$),
          \p2 = ($(9em,-1.2em)$) in
    [pattern={Lines[angle=45,distance=2pt]},pattern color=spkblue] (\p1) rectangle (\p2);
\draw[black](9em,0em) -- (1em,0em)  -- (1em,-1.2em) -- (9em,-1.2em);
\node[font=\scriptsize,text=black,inner sep=0pt,anchor=base] at ($(5em,-0.8em)$) {\contour{white}{Loose label}};

\draw[black, line width=0.5pt] (0em,-3.7em) -- (9.0em,-3.7em);
\path let \p1 = ($(3em,-2.5em)$),
          \p2 = ($(9em,-3.7em)$) in
    [pattern={Lines[angle=-45,distance=2pt]},pattern color=spkred] (\p1) rectangle (\p2);
\draw[black](9em,-2.5em) -- (3em,-2.5em) -- (3em,-3.7em) -- (9em,-3.7em);
\node[font=\scriptsize,text=black,inner sep=0pt,anchor=base] at ($(6em,-3.3em)$) {\contour{white}{Tight label}};
\node[draw=none,align=center,font=\scriptsize,inner sep=0] (t_1) at (3em, -1.9em) {$t_1'$};
\node[draw=none,align=center,font=\scriptsize,inner sep=0] (t_1) at (1em, 0.5em) {$t_1$};

\draw[black,dash pattern=on 2pt off 1pt,line width=0.5pt,color=gray] (3.0em,-3.7em) -- (3.0em,-4.5em);
\draw[black,dash pattern=on 2pt off 1pt,line width=0.5pt,color=gray] (1.0em,-1.2em) -- (1.0em,-4.5em);
\draw[<->, >={Latex[length=3pt,width=3pt]}, black, line width=0.4pt]
  (1em,-4.2em) -- node[midway, below, xshift=2.0em,font=\scriptsize, inner sep=4pt] {$\Delta_\text{onset}=t_1'-t_1$}
  (3em,-4.2em);

\end{tikzpicture}%
\hfill
    \includegraphics[width=0.32\linewidth]{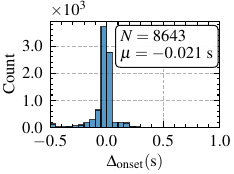}\hfill
    \includegraphics[width=0.32\linewidth]{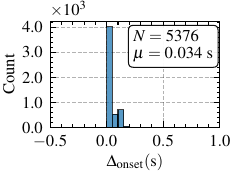}
\end{minipage}%
}
\hfill
\subfloat[Offset boundary shift\label{fig:offset_padding}]{%
\begin{minipage}[t]{\linewidth}
    \centering
    \begin{tikzpicture}[semithick,auto,
label/.style={
    draw=none,
    align=center,
    font=\footnotesize,
    inner sep=0,
    outer sep=0
},
]%

\definecolor{spkblue}{HTML}{0080B1}
\definecolor{spkred}{HTML}{E4002B}
\definecolor{spkgreen}{HTML}{06C755}

\draw[black, line width=0.5pt] (0em,-1.2em) -- (9.0em,-1.2em);
\path let \p1 = ($(0em,0em)$),
          \p2 = ($(8em,-1.2em)$) in
    [pattern={Lines[angle=45,distance=2pt]},pattern color=spkblue] (\p1) rectangle (\p2);
\draw[black](0em,0em) -- (8em,0em) -- (8em,-1.2em) -- (0em,-1.2em);
\node[font=\scriptsize,text=black,inner sep=0pt,anchor=base] at ($(4em,-0.8em)$) {\contour{white}{Loose label}};

\draw[black, line width=0.5pt] (0em,-3.7em) -- (9.0em,-3.7em);
\path let \p1 = ($(0em,-2.5em)$),
          \p2 = ($(6em,-3.7em)$) in
    [pattern={Lines[angle=-45,distance=2pt]},pattern color=spkred] (\p1) rectangle (\p2);
\draw[black](0em,-2.5em) -- (6em,-2.5em) -- (6em,-3.7em) -- (0em,-3.7em);
\node[font=\scriptsize,text=black,inner sep=0pt,anchor=base] at ($(3em,-3.3em)$) {\contour{white}{Tight label}};
\node[draw=none,align=center,font=\scriptsize,inner sep=0] (t_1) at (6em, -1.9em) {$t_2'$};
\node[draw=none,align=center,font=\scriptsize,inner sep=0] (t_1) at (8em, 0.5em) {$t_2$};

\draw[black,dash pattern=on 2pt off 1pt,line width=0.5pt,color=gray] (6.0em,-3.7em) -- (6.0em,-4.5em);
\draw[black,dash pattern=on 2pt off 1pt,line width=0.5pt,color=gray] (8.0em,-1.2em) -- (8.0em,-4.5em);
\draw[<->, >={Latex[length=3pt,width=3pt]}, black, line width=0.4pt]
  (6em,-4.2em) -- node[midway, below, xshift=-1.0em,font=\scriptsize, inner sep=4pt] {$\Delta_\text{offset}=t_2-t_2'$}
  (8em,-4.2em);
\end{tikzpicture}%
\hfill
    \includegraphics[width=0.32\linewidth]{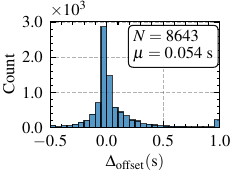}\hfill
    \includegraphics[width=0.32\linewidth]{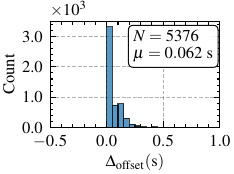}
\end{minipage}%
}
\caption{Amount of looseness caused by pause filling and onset/offset boundary shift. In each subfigure, the left panel illustrates pause filling or onset/offset boundary shift, while the middle and right panels show the corresponding histograms for AMI and AliMeeting, respectively. Values outside the displayed range are accumulated in the leftmost and rightmost bins, respectively.}
\label{fig:analyze_looseness}
\end{figure}
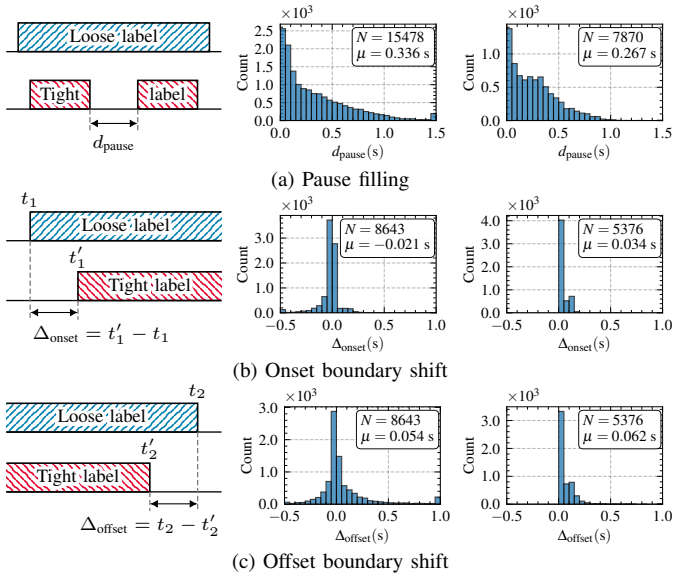

\subsubsection{Analysis}
We first analytically examine the extent of padding and pause filling using the validation sets of the AMI~\cite{carletta2007unleashing} and AliMeeting~\cite{yu2022m2met} corpora.
Although the method described in \autoref{sec:tight_boundary} does not require tight supervision, we chose these corpora for the analysis because tight annotations are available through forced alignment.
The total number of utterances is 8\,664 for AMI and 5\,378 for AliMeeting.
Of these utterances, 21 from AMI and 2 from AliMeeting are discarded due to alignment errors, and we therefore analyze the remaining 8\,643 and 5\,376 utterances, respectively.

\Autoref{fig:pause_filling} shows the distribution of the durations of pauses filled by loose annotations, as revealed by forced alignment.
The number of pauses is 15\,478 for AMI and 7\,870 for AliMeeting, with average durations of \SI{0.336}{\second} and \SI{0.267}{\second}, respectively.
Figures \ref{fig:onset_padding} and \ref{fig:offset_padding} show the distributions of the amount of shift at the onset and offset boundaries of loose annotations in AMI and AliMeeting.
Note that positive values indicate padding, whereas negative values indicate boundary truncation.
In contrast to pause filling, the average onset and offset boundary shifts are less than \SI{0.1}{\second}, indicating that boundary padding is much smaller than pause filling.

These observations can be attributed to the annotation process.
Annotators appear to perceive tight speech regions reasonably well, as implied by the small amount of onset and offset padding.
However, in ASR corpora, they tend to connect such regions by filling pauses to preserve semantic units, such as sentence-level segments.
This observation is also consistent with the finding of prior work that loose labels can be well reconstructed by closing operations without considering boundary padding~\cite{horiguchi2025can}.
Based on these findings, our label tightening focuses on undoing pause filling while leaving boundary padding unchanged to avoid over-tightening.
This is expected to prevent errors from increasing due to over-tightening at onset and offset before the model has sufficiently learned to suppress pause filling, which could otherwise misleadingly suggest that training has already converged.

\subsubsection{Method}
For a given speaker $s$, suppose that the interval from the $t_1$-th frame to the $t_2$-th frame forms a single loose segment, i.e., $\tilde{y}_{s,t}=1$ for $t_1\leq t\leq t_2$, while $\tilde{y}_{s,t_1-1}=\tilde{y}_{s,t_2+1}=0$.
Let $y_{\mathrm{tight},s,t}(\leq \tilde{y}_{s,t})$ be the element of tightened labels $Y_\mathrm{tight}$ in \autoref{algline:mask} in \autoref{algo:pseudo}.
Prior work attempted to prevent over-tightening by exploiting the fact that each loose segment should contain at least some speech (\autoref{algline:restore} in \autoref{algo:pseudo}).
Specifically, when the proportion of speech frames within a loose segment
\begin{equation}
\rho=\frac{\sum_{t=t_1}^{t_2}y_{\mathrm{tight},s,t}}{\sum_{t=t_1}^{t_2}\tilde{y}_{s,t}}
\end{equation}
fell below a predefined threshold, the original loose segment was used as the training target.
This threshold was set ad hoc to 0.5.
However, the appropriate value is clearly not common to all utterances.
For example, from the perspective of utterance duration, a short utterance is unlikely to contain pauses, and thus a larger threshold would be appropriate.
In contrast, a long utterance is more likely to contain pauses, and thus a smaller threshold would be preferable.

\begin{algorithm}[t]
    \caption{Loose boundary preservation}\label{algo:loose_preservation}
    \DontPrintSemicolon
    \SetKw{KwAnd}{and}
    \SetKwInOut{Input}{Input}
    \SetKwInOut{Output}{Output}
    \SetKwComment{Comment}{$\triangleright$\ }{}
    \Input{Start and end indices $(t_1, t_2)$ of a loose segment for speaker $s$, tightened label $Y_{\mathrm{tight}}=(y_{\mathrm{tight},s,t})_{t=t_1}^{t_2}$}
    \Output{Boundary-preserved tightened label $Y_{\mathrm{tight}}$}
    \BlankLine
    $t \leftarrow t_1$\label{algline:onset_preserve_start}\;
    \While(\Comment*[f]{Preserve loose onset}){$t \le t_2$ \KwAnd $y_{\mathrm{tight},s,t}=0$}{
        $y_{\mathrm{tight},s,t}\leftarrow 1$\;
        $t \leftarrow t+1$\label{algline:onset_preserve_end}\;
    }
    $t \leftarrow t_2$\label{algline:offset_preserve_start}\;
    \While(\Comment*[f]{Preserve loose offset}){$t \ge t_1$ \KwAnd $y_{\mathrm{tight},s,t}=0$}{
        $y_{\mathrm{tight},s,t}\leftarrow 1$\;
        $t \leftarrow t-1$\label{algline:offset_preserve_end}\;
    }
\end{algorithm}

In this work, we always preserve the onset and offset boundaries of each loose segment, and therefore do not attempt to reduce boundary padding.
This can be achieved simply by using \autoref{algo:loose_preservation}.
On the onset side, we start from $t_1$, increment the time index, and set all frames to 1 until reaching the onset of the first estimated tight segment (lines~\ref{algline:onset_preserve_start}--\ref{algline:onset_preserve_end}).
Similarly, on the offset side, we start from $t_2$, decrement the time index, and set all frames to 1 until reaching the offset of the last estimated tight segment (lines~\ref{algline:offset_preserve_start}--\ref{algline:offset_preserve_end}).
If no tight segment exists inside the loose segment, the original loose segment is fully restored.
Therefore, this procedure also satisfies the condition intended by the prior method, namely that each loose segment should contain speech.

\subsection{Causal and anticausal models with a burn-in phase}\label{sec:burn_in}
\subsubsection{Analysis}

While pseudo-labels are generated using the outputs of causal and anticausal models, these models' performance is generally lower than that of non-causal models.
We analyze the conditions under which these errors occur most frequently.

The models used in this paper process \SI{10}{\second} chunks, and \autoref{fig:without_burnin} shows the position-wise DER of the causal and anticausal models within each chunk, where each position corresponds to a \SI{1}{\second} subchunk.
The results show that missed detections increase particularly near the beginning of the sequence, i.e., \qtyrange[range-phrase=--,range-units=single]{0}{1}{\second} for the causal model and \qtyrange[range-phrase=--,range-units=single]{9}{10}{\second} for the anticausal model.
This degradation can be attributed to a cold-start problem, where it must extract features and detect speakers from very limited samples at the beginning.

\subsubsection{Method}
We address the issue by introducing a burn-in phase~\cite{bianchi2017overview,beintema2021nonlinear,schiller2026tuning}, which allows unidirectional models to accumulate sufficient context before producing the outputs used for diarization.
In the case of a causal model $\causal{f}(\cdot)$, the model takes an extra input $X'=[\vect{x}_{-L'+1},\dots,\vect{x}_0]$ of length $L'$, corresponding to the samples preceding the original input $X$:
\begin{equation}
[\underbrace{\vect{q}_{-T'+1},\dots,\vect{q}_{0}}_{\text{associated with }X'},\underbrace{\vect{q}_1,\dots,\vect{q}_T}_{\text{associated with }X}]=\causal{f}\bigl([X'\:\: X]\bigr).
\end{equation}
Note that $T'$ is the number of output frames associated with $X'$.
Similarly, in the case of an anticausal model $\anticausal{f}(\cdot)$, the model takes an extra input $X'=[\vect{x}_{L+1},\dots,\vect{x}_{L+L'}]$ of length $L$, corresponding to the samples following the original input $X$:
\begin{equation}
[\underbrace{\vect{q}_1,\dots,\vect{q}_T}_{\text{associated with }X},\underbrace{\vect{q}_{T+1},\dots,\vect{q}_{T+T'}}_{\text{associated with }X'}]=\anticausal{f}\bigl([X\:\: X']\bigr).
\end{equation}
The outputs associated with the extra inputs $X'$ are used only for burn-in; they are excluded from the loss computation during training and discarded during inference.
In this paper, we set the burn-in phase to \SI{1}{\second}.
These extra inputs allow unidirectional models to access sufficient context from the beginning of the input sequence.

\subsection{Non-causal-model-aware co-training}
\label{sec:non_causal_aware}

\begin{algorithm}[t]
    \caption{Co-training of causal and anticausal models, with parallel training of the final non-causal model}\label{algo:cotrain_noncausal}
    \DontPrintSemicolon
    \SetKwInOut{Input}{Input}
    \SetKwInOut{Output}{Output}
    \SetKwComment{Comment}{$\triangleright$\ }{}
    \Input{Training dataset $\mathcal{D}=\{(X_i,\tilde{Y}_i)\}_{i=1}^N$, causal model $\causal{f}$ with parameters $\causal{\theta}$, anticausal model $\anticausal{f}$ with parameters $\anticausal{\theta}$, non-causal model $f$ with parameters $\theta$}
    \Output{Updated non-causal model's parameters $\theta$}
    \BlankLine

    \While{non-causal model has not converged\label{algline:monitor}}{
        Sample minibatch $(\mathbf{X},\tilde{\mathbf{Y}})$ from $\mathcal{D}$\;\label{algline:start_conventional}
        \tcp{(i) Pseudo-label generation}
        $\causal{\mathbf{Q}}\leftarrow \causal{f}(\mathbf{X};\causal{\theta})$\label{algline:pseudo_start}\label{algline:forward_causal}\Comment*[r]{Forward causal}
        $\anticausal{\mathbf{Q}}\leftarrow\anticausal{f}(\mathbf{X};\anticausal{\theta})$\Comment*[r]{Forward anticausal}
        $\causal{\mathbf{P}},\anticausal{\mathbf{P}}\leftarrow\mathsf{Q2P}(\causal{\mathbf{Q}}),\mathsf{Q2P}(\anticausal{\mathbf{Q}})$\Comment*[r]{Eq. \autoref{eq:q2p}}
        $\mathbf{Y}_\mathrm{tight}\leftarrow \mathsf{Tighten}\bigl(\tilde{\mathbf{Y}},\causal{\mathbf{P}},\anticausal{\mathbf{P}}\bigr)$\Comment*[r]{Alg.~\ref{algo:pseudo}\&\autoref{sec:pause_fill}\&\ref{sec:burn_in}}\label{algline:pseudo_end}
        \tcp{(ii) Update causal and anticausal models}
        Update $\causal{\theta}$ on $\ell(\causal{\mathbf{Q}},\mathbf{Y}_\mathrm{tight})$\label{algline:update_causal}\Comment*[r]{Backward causal}
        Update $\anticausal{\theta}$ on $\ell(\anticausal{\mathbf{Q}},\mathbf{Y}_\mathrm{tight})$ \label{algline:update_anticausal}\Comment*[r]{Backward anticausal}\label{algline:end_conventional}
        \tcp{(iii) Update non-causal model}
        $\mathbf{Q}\leftarrow f(\mathbf{X};\theta)$\Comment*[r]{Forward non-causal}\label{algline:forward_noncausal}
        Update $\theta$ on $\ell(\mathbf{Q},\mathbf{Y}_\mathrm{tight})$ \label{algline:update_student_end}\Comment*[r]{Backward non-causal}\label{algline:backward_noncausal}
    }
\end{algorithm}

In the pipeline of the prior study, as described in \autoref{sec:pseudo_labeling}, causal and anticausal models are first co-trained in \textbf{Step 2} and then used to generate pseudo-labels for training the final non-causal model in \textbf{Step 3}.
However, pseudo-labels from the best causal and anticausal models do not necessarily maximize the performance of the final non-causal model, potentially yielding outputs that are either over-tightened or insufficiently tight.
This concern is consistent with findings in knowledge distillation, where the best-performing teacher model does not always lead to the best student model~\cite{cho2019efficacy,park2021learning,stanton2021does,dong2024toward}, and with pseudo-labeling studies showing that noisy or overconfident pseudo-labels can induce confirmation bias and noisy training~\cite{arazo2020pseudo,rizve2021in}.

To address this issue, the proposed pipeline makes the co-training process aware of the final non-causal model by training it in parallel and using its validation performance for early stopping.
We replace the procedure described in \autoref{sec:pseudo_labeling} with the following procedure for training a non-causal model capable of producing tight outputs:
\begin{itemize}
\item \textbf{Step 1:} Train causal, anticausal, and \underline{non-causal} models independently using loose labels.
\item \textbf{Step 2:} Starting from their respective pretrained weights, co-train the causal and anticausal models while training the \underline{non-causal} model in parallel. At each training iteration, pseudo-labels are generated on the fly from the outputs of the causal and anticausal models and are used to update all three models.
\end{itemize}

\autoref{algo:cotrain_noncausal} shows the detailed co-training procedure.
The bold-faced symbols, e.g., $\causal{\mathbf{Q}}$, denote batched versions of the corresponding original symbols, e.g., $\causal{Q}$.
As in the prior method, each iteration starts with a minibatch (\autoref{algline:start_conventional}), generates tight pseudo-labels using the causal and anticausal models (lines~\ref{algline:pseudo_start}--\ref{algline:pseudo_end}), and updates these models using the generated pseudo-labels (lines~\ref{algline:update_causal}--\ref{algline:update_anticausal}).
The key difference is that the proposed method also updates the non-causal model, initialized from loose-label training, using the pseudo-labels (lines~\ref{algline:forward_noncausal}--\ref{algline:backward_noncausal}).
This avoids introducing a separate learning schedule for training the final model from scratch.

In addition, the stopping criterion is based on the validation performance of the non-causal model (\autoref{algline:monitor}).
In the prior method, early stopping was determined by the convergence of the causal and anticausal models.
However, pseudo-labels generated using the best causal and anticausal models do not necessarily maximize the performance of the resulting non-causal model.
Therefore, the proposed pipeline directly monitors the final non-causal model during the training loop and selects the training point that is better suited to the model used for final inference.

\section{Experimental Setup}
\subsection{Diarization pipeline}
Our experiments are based on the EEND-vector clustering framework, which performs local speaker diarization using a sliding window and then aggregates the window-level results by clustering speaker embeddings~\cite{kinoshita2021integrating}.
The proposed method is applied to the local speaker diarization model used in the former step.
In the latter step, we extracted speaker embeddings using the ResNet34 extractor trained with VoxCeleb~\cite{nagrani2020voxceleb}\footnote{\fontsize{7.5pt}{8.5pt}\selectfont\url{https://huggingface.co/Wespeaker/wespeaker-voxceleb-resnet34-LM}}, and then clustered them via VBx clustering~\cite{landini2022bayesian,palka2026vbx} using the corresponding PLDA model\footnote{\fontsize{7.5pt}{8.5pt}\selectfont{\url{https://huggingface.co/BUT-FIT/diarizen-wavlm-base-s80-md/tree/main/plda}}}.
The VBx clustering parameters, i.e., $F_A$ and $F_B$, are tuned on the validation set of each corpus using the pyannote.audio toolkit~\cite{bredin2020pyannote}.

For local speaker diarization, we followed the prior study~\cite{horiguchi2026tight} and used the architecture based on a multi-speaker embedding extractor~\cite{horiguchi2025pretraining}, as described in detail in \autoref{sec:architecture}.
The non-causal, causal, and anticausal ReDimNet encoders were first pretrained on VoxCeleb-based simulated mixtures following the same pretraining procedure~\cite{horiguchi2025pretraining}.
The backend, consisting of an LSTM and a linear layer, was then attached to the encoder and trained as a diarization model using the compound set. 
The model and training configurations follow the corresponding prior studies and are summarized in \autoref{tbl:training_configurations}.

\begin{table}
\caption{Detailed model and training configurations of local diarization of the EEND-VC pipeline}
\label{tbl:training_configurations}
\begin{tabularx}{\linewidth}{@{}l>{\raggedright\arraybackslash}X@{}}
\toprule
Item&Configuration\\\midrule
Sliding window&\SI{10}{\second} width / \SI{1}{\second} shift\\\customdashline{1-2}
Max. \# of speakers&4 speakers per chunk ($S=4$)\\
&2 speakers per frame ($M=2$)\\\customdashline{1-2}
Optimizer&Adam~\cite{kingma2015adam}\\\customdashline{1-2}
Learning rate scheduler&\textbf{Steps 1\&3} in \autoref{sec:pseudo_labeling} and \textbf{Step 1} in \autoref{sec:non_causal_aware}: 1k-step linear warmup to 0.001; exponential decay by 0.8 every 6k steps\\
&\textbf{Step 2} in \autoref{sec:pseudo_labeling} and \textbf{Step 2} in \autoref{sec:non_causal_aware}: Fixed to 0.0001\\
\bottomrule
\end{tabularx}
\end{table}

\subsection{Dataset}
\begin{table}[t]
\centering
\caption{Statistics of the dataset used in the experiments}\label{tbl:dataset}
\begin{tabular}{@{}llrrr@{}}
\toprule
&&\multicolumn{3}{c@{}}{Duration (hours)}\\\cmidrule(l){3-5}
Annotation style&Corpus&Train&Val&Test\\\midrule
Loosely annotated&AMI-MHM&80.7&9.7&9.1\\
&AMI-SDM&79.7&9.7&9.1\\
&AliMeeting&111.4&4.2&10.8\\\customdashline{1-5}
Tightly annotated&MSDWild&64.1&2.0&9.9\\
&VoxConverse&18.3&2.0&43.5\\\midrule
Total & Compound set&352.3&27.6&82.3\\
\bottomrule
\end{tabular}
\end{table}

A summary of the datasets used in the experiments is provided in \autoref{tbl:dataset}.
We used the compound set of five corpora for model training: AMI mixed headset microphones (AMI-MHM) and single distant microphones (AMI-SDM)~\cite{carletta2007unleashing}, AliMeeting~\cite{yu2022m2met}, MSDWild few-talker set (MSD)~\cite{liu2022msdwild}, and VoxConverse (VoxC)~\cite{chung2020spot}.
Note that AMI-MHM, AMI-SDM, and AliMeeting are ASR corpora, so they have loosely annotated labels, whereas MSD and VC are diarization corpora with tightly annotated labels.

Although the proposed method does not require tight labels for training on the training set, validation and evaluation still require tight labels.
We therefore used AMI and AliMeeting, for which publicly available tight labels obtained via forced alignment are available.\footnote{\url{https://github.com/nttcslab-sp/diar-forced-alignment}} 
Specifically, we assume that these tight labels are available on the validation sets for monitoring training convergence, which is common practice in noisy-label training~\cite{ren2018learning,shu2019meta}.
We then evaluate each system on the test portion of each corpus in terms of diarization error rate (DER) without collar tolerance, using tight labels as references.

\section{Results}

\subsection{Speaker diarization}
\begin{table*}[t]
\caption{DER (\si{\percent}) on each corpus. Tight labels are always used as reference; for AMI-MHM, AMI-SDM, and AliMeeting, forced-aligned labels are used. Pause-fill, Burn-in, and Non-causal denote pause-filling-focused tightening (\autoref{sec:pause_fill}), introduction of a burn-in phase (\autoref{sec:burn_in}), and non-causal-model-aware co-training (\autoref{sec:non_causal_aware}), respectively. Values in parentheses indicate the error breakdown: missed detection, false alarm, and speaker confusion, respectively.}
\renewrobustcmd{\bfseries}{\fontseries{b}\selectfont}
\renewrobustcmd{\boldmath}{}
\newrobustcmd{\B}{\bfseries}
\label{tbl:results}
\centering
\sisetup{detect-weight,mode=text}
\setlength{\tabcolsep}{3pt}
\begin{tabular}{@{}l@{}ccc*{3}{S[table-format=2.2]@{\;\scriptsize(}r@{\scriptsize\,/\,}r@{\scriptsize\,/\,}r@{{\scriptsize)}\hspace{2\tabcolsep}}}*{2}{S[table-format=2.2]}@{}}
\toprule
&\multirow{2.92}{*}{\makecell{Pause-\\fill}}&\multirow{2.92}{*}{\makecell{Burn-\\in}}&\multirow{2.92}{*}{\makecell{Non-\\causal}}&\multicolumn{12}{c}{Original annotation: Loose}&\multicolumn{2}{c}{Tight}\\\cmidrule(l{\tabcolsep}r{2\tabcolsep}){5-16}\cmidrule{17-18}
Label source for AMI/AliMeeting&&&&\multicolumn{4}{c@{\hspace{2\tabcolsep}}}{AMI-MHM}&\multicolumn{4}{@{}c@{\hspace{2\tabcolsep}}}{AMI-SDM}&\multicolumn{4}{@{}c@{\hspace{2\tabcolsep}}}{AliMeeting}&{MSD}&{VoxC}\\\midrule
($\mathtt{L1}$) \textit{Original loose annotations}&&&&\textit{32.10}&\scriptsize \textit{1.64}&\scriptsize \textit{29.98}&\scriptsize \textit{0.48}&\textit{32.10}&\scriptsize \textit{1.64}&\scriptsize \textit{29.98}&\scriptsize \textit{0.48}&\textit{16.90}&\scriptsize \textit{0.00}&\scriptsize \textit{16.90}&\scriptsize \textit{0.00}&\textit{N/A}&\textit{N/A}\\\midrule
\multicolumn{11}{@{}l}{\textbf{Conventional method}}\\
($\mathtt{C1}$) Trained on loose label (original)&&&&33.97&\scriptsize 3.10&\scriptsize 26.00&\scriptsize 4.87&36.34 &\scriptsize 3.99&\scriptsize 25.54&\scriptsize 6.81&26.54&\scriptsize 4.54&\scriptsize 14.18&\scriptsize 7.81&21.35&9.91\\
($\mathtt{C2}$) Trained on tight label (forced-aligned)&&&&14.03&\scriptsize 5.88&\scriptsize 3.88&\scriptsize 4.27&17.05&\scriptsize 7.74&\scriptsize 4.27&\scriptsize 5.05&19.79&\scriptsize 7.96&\scriptsize 4.76&\scriptsize 7.07&21.35&10.75\\
\customdashline{1-18}
($\mathtt{C3}$) Trained on tight pseudo-label (VAD)~\cite{horiguchi2026tight}&&&&18.99&\scriptsize 7.22&\scriptsize 7.38&\scriptsize 4.39&23.75&\scriptsize 9.04&\scriptsize 8.56&\scriptsize 6.15&23.36&\scriptsize 9.15&\scriptsize 6.68&\scriptsize 7.54&21.47&10.08\\
($\mathtt{C4}$) Trained on tight pseudo-label (SC)~\cite{horiguchi2026tight}&&&&18.28&\scriptsize 7.78&\scriptsize 5.80&\scriptsize 4.70&21.56&\scriptsize 9.67&\scriptsize 5.98&\scriptsize 5.91&22.90&\scriptsize 11.20&\scriptsize 4.92&\scriptsize 6.79&21.96&10.19\\\midrule
\multicolumn{11}{@{}l}{\textbf{Proposed method}}\\
($\mathtt{P1}$) Trained on tight pseudo-label (SC)&$\checkmark$&&&16.49&\scriptsize 7.46&\scriptsize 4.92&\scriptsize 4.11&20.05&\scriptsize 9.77&\scriptsize 4.90&\scriptsize 5.38&21.22&\scriptsize 7.59&\scriptsize 7.27&\scriptsize 6.36&21.58&10.33\\
($\mathtt{P2}$) Trained on tight pseudo-label (SC)&$\checkmark$&$\checkmark$&&16.44&\scriptsize 6.86&\scriptsize 5.51&\scriptsize 4.07&19.78&\scriptsize 9.25&\scriptsize 5.39&\scriptsize 5.15&\B 21.02&\scriptsize 6.77&\scriptsize 8.34&\scriptsize 5.91&21.26&10.15\\
($\mathtt{P3}$) Trained on tight pseudo-label (SC)&$\checkmark$&\checkmark&$\checkmark$&\B 16.40&\scriptsize 6.68&\scriptsize 5.60&\scriptsize 4.13&\B 19.63&\scriptsize 8.64&\scriptsize 5.68&\scriptsize 5.32&21.38&\scriptsize 6.73&\scriptsize 8.39&\scriptsize 6.29&21.24&10.22\\
\bottomrule
\end{tabular}%
\end{table*}

\autoref{tbl:results} reports the DERs of conventional and proposed approaches on each corpus.
Because speaker confusion depends in part on the accuracy of clustering across chunks, we mainly focus our discussion on missed detection and false alarms.
First, $\mathtt{L1}$ shows the DER of the loose labels themselves.
AMI has approximately twice the DER of AliMeeting, which, together with the analysis in \autoref{sec:pause_fill}, suggests that AMI contains roughly twice as much pause filling.

When evaluated against forced-aligned tight labels, $\mathtt{C1}$, trained with loose labels, performed substantially worse than the ideal tight-label topline $\mathtt{C2}$ on corpora with originally loose annotations.
This degradation was mainly reflected in high false-alarm rates, indicating that the model produced loose speech segments.
This observation is consistent with prior studies~\cite{horiguchi2025can,horiguchi2026tight}.
The pseudo-labeling methods in the prior study~\cite{horiguchi2026tight}, shown in $\mathtt{C3}$ and $\mathtt{C4}$, achieved performance close to $\mathtt{C2}$ without tight supervision, although a substantial gap still remained.
Although the SC-based method ($\mathtt{C4}$) outperformed the VAD-based method ($\mathtt{C3}$) in DER, its increased missed detection and speaker confusion have been reported to affect downstream tasks negatively~\cite{horiguchi2026tight}.
From $\mathtt{C2}$ to $\mathtt{C4}$, missed detection and false alarm increased to a similar extent on AMI, whereas the degradation on AliMeeting was dominated by missed detection.
This indicates that over-tightening and insufficient tightening occur to a similar extent on AMI, while over-tightening is dominant on AliMeeting.
As discussed in \autoref{sec:pause_fill}, this can be attributed to the fixed threshold used in loose segment restoration, despite the utterance-dependent appropriate threshold.
Specifically, since AMI contains more pause filling, a smaller threshold can be used.
In contrast, since AliMeeting contains less pause filling, using a small threshold tends to make the model learn overly tight outputs.

\begin{table*}[t]
\renewrobustcmd{\bfseries}{\fontseries{b}\selectfont}
\renewrobustcmd{\boldmath}{}
\newrobustcmd{\B}{\bfseries}
\setlength{\tabcolsep}{3.5pt}
\sisetup{detect-weight,detect-shape,mode=text}
\begin{minipage}[t]{0.37\linewidth}
\caption{Average onset/offset boundary shifts~(\si{\second}) relative to loose annotations. Positive values indicate tighter boundaries (see \autoref{fig:analyze_looseness} left).}
\label{tbl:boundary_error}
\centering
\scalebox{0.94}{%
\begin{tabular}{@{}lS[table-format=-1.3]@{\;/\;}S[table-format=-1.3]S[table-format=-1.3]@{\;/\;}S[table-format=-1.3]S[table-format=-1.3]@{\;/\;}S[table-format=-1.3]@{}}
\toprule
&\multicolumn{2}{c}{AMI-MHM}&\multicolumn{2}{c}{AMI-SDM}&\multicolumn{2}{c@{}}{AliMeeting}\\\midrule
($\mathtt{L1}$)&\itshape 0.000 & \itshape 0.000 & \itshape 0.000 & \itshape 0.000 & \itshape 0.000 & \itshape 0.000\\\midrule
($\mathtt{C1}$)&-0.142 & -0.160 & -0.160 & -0.179 & -0.272 & -0.393\\
($\mathtt{C2}$)&0.048 & 0.144 & 0.068 & 0.178 & 0.099 & 0.118\\\customdashline{1-7}
($\mathtt{C3}$)&0.063 & -0.079 & 0.067 & 0.105 & 0.150 & 0.082\\
($\mathtt{C4}$)&0.079 & 0.122 & 0.105 & 0.150 & 0.162 & 0.098\\\midrule
($\mathtt{P1}$)&0.072 & 0.107 & 0.088 & 0.143 & 0.007 & -0.008\\
($\mathtt{P2}$)&0.059 & 0.104 & 0.083 & 0.141 & -0.048 & -0.073\\
($\mathtt{P3}$)&0.064 & 0.103 & 0.080 & 0.139 & -0.050 & -0.069\\
\bottomrule
\end{tabular}%
}
\end{minipage}
\hfill
\begin{minipage}[t]{0.61\linewidth}
\centering
\caption{tcpWER (\si{\percent}) and ORC-WER (\si{\percent}) on AMI. Values in parentheses indicate the error breakdown: deletion, insertion, and substitution.}
\label{tbl:asr}
\scalebox{0.94}{%
\begin{tabular}{@{}lS[table-format=2.2]@{\;\scriptsize(}r@{\scriptsize\,/\,}r@{\scriptsize\,/\,}r@{{\scriptsize)}\hspace{2\tabcolsep}}S[table-format=2.2]@{\;\scriptsize(}r@{\scriptsize\,/\,}r@{\scriptsize\,/\,}r@{{\scriptsize)}\hspace{2\tabcolsep}}S[table-format=2.2]@{\;\scriptsize(}r@{\scriptsize\,/\,}r@{\scriptsize\,/\,}r@{{\scriptsize)}\hspace{2\tabcolsep}}S[table-format=2.2]@{\;\scriptsize(}r@{\scriptsize\,/\,}r@{\scriptsize\,/\,}r@{{\scriptsize)}}@{}}
\toprule
&\multicolumn{8}{c}{Multi-channel ASR (GSS\,+\,Whisper large-v3)}&\multicolumn{8}{c}{Single-channel ASR (SE-DiCoW)}\\\cmidrule(l{\tabcolsep}r{2\tabcolsep}){2-9}\cmidrule{10-17}
&\multicolumn{4}{c@{\hspace{2\tabcolsep}}}{tcpWER}&\multicolumn{4}{@{}c@{\hspace{2\tabcolsep}}}{ORC-WER}&\multicolumn{4}{@{}c@{\hspace{2\tabcolsep}}}{tcpWER}&\multicolumn{4}{c}{ORC-WER}\\\midrule
($\mathtt{C1}$)&26.75&\scriptsize 17.63&\scriptsize 3.25&\scriptsize 5.87&23.27&\scriptsize 15.37&\scriptsize 0.99&\scriptsize 6.91&24.32&\scriptsize 10.57&\scriptsize 6.91&\scriptsize 6.84&18.24&\scriptsize 6.93&\scriptsize 3.27&\scriptsize 8.04\\
($\mathtt{C2}$)&24.65&\scriptsize 16.34&\scriptsize 2.73&\scriptsize 5.59&21.61&\scriptsize 14.68&\scriptsize 1.07&\scriptsize 5.86&22.30&\scriptsize 9.67&\scriptsize 5.60&\scriptsize 7.03&17.41&\scriptsize 6.68&\scriptsize 2.60&\scriptsize 8.13\\\customdashline{1-17}
($\mathtt{C3}$)&26.44&\scriptsize 17.66&\scriptsize 3.19&\scriptsize 5.60&22.82&\scriptsize 15.58&\scriptsize 1.11&\scriptsize 6.12&23.70&\scriptsize 10.32&\scriptsize 6.46&\scriptsize 6.92&17.73&\scriptsize 6.73&\scriptsize 2.87&\scriptsize 8.13\\
($\mathtt{C4}$)&26.46&\scriptsize 17.80&\scriptsize 3.17&\scriptsize 5.48&22.74&\scriptsize 15.61&\scriptsize 0.98&\scriptsize 6.16&23.24&\scriptsize 10.41&\scriptsize 5.92&\scriptsize 6.91&17.34&\scriptsize 6.85&\scriptsize 2.33&\scriptsize 8.18\\\midrule
($\mathtt{P1}$)&25.00&\scriptsize 16.92&\scriptsize 2.72&\scriptsize 5.36&22.42&\scriptsize 15.20&\scriptsize 1.00&\scriptsize 6.22&22.65&\scriptsize 9.79&\scriptsize 5.83&\scriptsize 7.02&17.79&\scriptsize 6.77&\scriptsize 2.81&\scriptsize 8.21\\
($\mathtt{P2}$)&\B 24.72&\scriptsize 16.64&\scriptsize 2.71&\scriptsize 5.38&22.15&\scriptsize 14.97&\scriptsize 1.05&\scriptsize 6.13&21.62&\scriptsize 9.33&\scriptsize 5.17&\scriptsize 7.12&16.66&\scriptsize 6.24&\scriptsize 2.08&\scriptsize 8.34\\
($\mathtt{P3}$)&24.79&\scriptsize 16.59&\scriptsize 2.78&\scriptsize 5.42&\B 22.07&\scriptsize 14.91&\scriptsize 1.10&\scriptsize 6.06&\B 21.47&\scriptsize 9.27&\scriptsize 5.13&\scriptsize 7.07&\B 16.62&\scriptsize 6.36&\scriptsize 2.21&\scriptsize 8.05\\
\bottomrule
\end{tabular}%
}
\end{minipage}
\end{table*}

Focusing on pause filling, the proposed method ($\mathtt{P1}$) improved the DER on average and substantially mitigated this asymmetry.
On AMI, the method reduced false alarms while maintaining a similar level of missed detection.
In contrast, on AliMeeting, it substantially reduced missed detection with a smaller accompanying increase in false alarms, indicating that over-tightening is effectively alleviated.
This is also evident from the average onset and offset boundary shifts shown in \autoref{tbl:boundary_error}.
Compared with $\mathtt{C2}$, the $\mathtt{C4}$ system showed a larger amount of tightening at onsets, indicating over-tightening.
In contrast, all methods with pause-filling-focused tightening, i.e., $\mathtt{P1}$--$\mathtt{P3}$, mitigated such excessive boundary shifts compared with $\mathtt{C4}$, especially on AliMeeting.
This behavior is consistent with the design of the proposed tightening, which suppresses pause filling without aggressively shifting loose onset and offset boundaries.

\begin{figure}[t]
\centering
\includegraphics[width=0.7\linewidth]{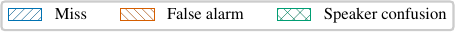}\\\vspace{0.3em}
\subfloat[Conventional (without burn-in)\label{fig:without_burnin}]{%
\includegraphics[width=0.48\linewidth]{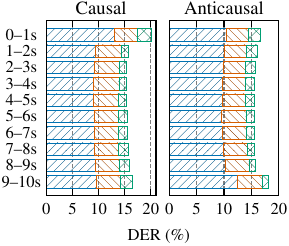}%
}
\hfill
\subfloat[Proposed (with burn-in)\label{fig:with_burnin}]{%
\includegraphics[width=0.48\linewidth]{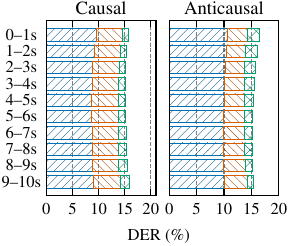}%
}
\caption{Position-wise DER and its breakdown of the causal and anticausal models without and with a burn-in phase, within \SI{10}{\second} chunks. DER is computed separately for each 1-\si{\second} subchunk according to its relative position in the chunk, e.g., \numrange[range-phrase={--}]{0}{1}, \numrange[range-phrase={--}]{1}{2}, \dots, and \qtyrange[range-phrase=--,range-units=single]{9}{10}{\second}. The optimal label permutation is recomputed for each subchunk.}
\label{fig:burnin}
\end{figure}

Introducing the burn-in phase to the causal and anticausal models further reduced missed detection, at the cost of a comparable increase in false alarms ($\mathtt{P2}$).
To examine the source of this effect, \autoref{fig:burnin} shows the position-wise DER within each chunk with and without the burn-in phase.
Without the burn-in phase, misses increased by 2--3 percentage points near the beginning of the context available to each unidirectional model, i.e., \qtyrange[range-phrase=--,range-units=single]{0}{1}{\second} for the causal model and \qtyrange[range-phrase=--,range-units=single]{9}{10}{\second} for the anticausal model.
With the burn-in phase, this degradation was substantially mitigated, resulting in approximately uniform DER across positions within chunks.

Finally, $\mathtt{P3}$ in \autoref{tbl:results} shows the result of non-causal-model-aware co-training, where the final non-causal model was trained in parallel with the co-training of the causal and anticausal models.
Although the improvement is modest because this method does not directly improve the pseudo-label generation process itself, it reduced missed detection at the cost of increased false alarms.

\subsection{Multi-talker ASR}
\subsubsection{Evaluation protocol}
We evaluated the impact of diarization results on downstream multi-talker ASR.
To cover both multi-channel and single-channel settings, we used two systems that generate speaker-wise transcriptions conditioned on diarization results.

As a multi-channel ASR pipeline, we used a cascaded system that performs source separation followed by ASR.
Specifically, it first applies guided source separation (GSS)~\cite{boeddeker2018front,raj2023gpu} to obtain utterance-wise separated signals and then transcribes them using Whisper large-v3~\cite{radford2023robust}.
Following prior work~\cite{horiguchi2025can}, the diarization results were directly used in the time-frequency mask estimation step of GSS, while morphological closing was applied before constructing the beamformer and performing ASR.
For evaluation, we used the first 8-channel circular microphone array recordings of the AMI corpus.
The first channel of this microphone array corresponds to the AMI-SDM recordings used in the diarization experiments, and we used the diarization outputs estimated from AMI-SDM.

The second method is self-enrolled diarization-conditioned Whisper (SE-DiCoW)~\cite{polok2026sedicow}, which is an end-to-end single-channel multi-talker ASR model.
This model takes multi-talker recordings and the corresponding diarization results as inputs.
For evaluation, we used AMI-SDM together with the diarization outputs estimated from it.

We used the time-constrained minimum-permutation word error rate (tcpWER)~\cite{von2023meeteval} with \SI{5}{\second} collar as an evaluation metric.
In diarization-based ASR, speaker confusion errors in diarization cause transcriptions to be assigned to the wrong speaker, which results in paired deletion and insertion errors in tcpWER.
Therefore, to assess ASR performance with reduced sensitivity to such speaker confusion errors, we also report reference combination word error rate (ORC-WER)~\cite{sklyar2022multi}.
Both metrics were computed using the MeetEval toolkit after applying the CHiME-8 text normalization~\cite{von2023meeteval}.

\subsubsection{Results}
In the multi-channel ASR results shown in \autoref{tbl:asr}, the conventional pseudo-tightening-based systems, $\mathtt{C3}$ and $\mathtt{C4}$, achieved better tcpWER and ORC-WER than the loose-label-based system ($\mathtt{C1}$).
However, the improvement was smaller than that observed in DER, and most of the gap from the system trained with tight labels ($\mathtt{C2}$) was attributed to deletion errors.
One possible reason is that missed detections in diarization are unrecoverable in the cascaded system.
In contrast, the systems using the proposed methods ($\mathtt{P1}$--$\mathtt{P3}$) substantially reduced deletion errors, achieving ASR performance closer to that of the $\mathtt{C2}$ system.

For single-channel ASR, even the over-tightened systems, $\mathtt{C3}$ and $\mathtt{C4}$, achieved results closer to $\mathtt{C2}$ than to $\mathtt{C1}$, especially in terms of ORC-WER.
This can be explained by the fact that SE-DiCoW incorporates augmentation that simulates label errors during training, making it robust to diarization errors.
Even under this setting, the systems using the proposed methods, especially $\mathtt{P2}$ and $\mathtt{P3}$, outperformed conventional pseudo-label-based systems ($\mathtt{C3}$ and $\mathtt{C4}$), and even surpassed the $\mathtt{C2}$ system trained with tight labels.
A possible reason is that SE-DiCoW is trained using the original loose labels, and therefore fully tight diarization outputs may be out of domain for SE-DiCoW.
Using DiCoW trained with tight labels~\cite{polok2026mind} may further exploit the potential of tightly predicted diarization results, as observed in multi-channel ASR, which is left for future work.

\section{Conclusion}
In this paper, we investigated several factors that cause over-tightening in pseudo-tight-label generation based on causal--anticausal consistency for tight-boundary speaker diarization.
We addressed these factors by preserving loose boundaries during pseudo-label generation, introducing a burn-in phase for causal and anticausal models, and non-causal-model-aware co-training.
Experimental results demonstrated improvements in both speaker diarization and downstream multi-talker ASR performance.

\clearpage
\section{Generative AI Use Disclosure}
ChatGPT was used for English polishing. All the technical content was developed and verified by the authors.
\IEEEtriggeratref{50}
\bibliographystyle{IEEEbib}
\bibliography{mybib}

\end{document}